\documentclass{preprint}

\usepackage{graphicx}
\usepackage[dvipsnames]{xcolor}
\definecolor{neonMagenta}{rgb}{1,0,1}
\definecolor{neonCyan}{rgb}{0,0.7,0.7}
\definecolor{neonOrange}{rgb}{1,0.5,0}
\definecolor{neonPurple}{rgb}{0.5,0.25,1}
\definecolor{soberGreen}{rgb}{0,0.5,0}
\definecolor{sky}{rgb}{0.225,0.57,0.8}
\definecolor{maroon}{rgb}{0.7059,0.0157,0.149}
\usepackage{newtxtext}
\usepackage{newtxmath}
\usepackage[round,authoryear]{natbib}
\usepackage{hyperref}
\hypersetup{
    colorlinks = true,
    urlcolor   = blue,
    linkcolor  = black,
    pdftitle   = {(Why) Jets quantise trapped acoustic modes},
    pdfauthor  = {Brandon Yeung, Michael N. Stavropoulos, Vincent Jaunet, Peter Jordan, Oliver T. Schmidt},
    citecolor  = black,
}
\usepackage{physics}
\usepackage[retainorgcmds]{IEEEtrantools}
\usepackage{tikz}
\usetikzlibrary{arrows.meta,backgrounds}

\newcommand{\beq}{\begin{equation}}
\newcommand{\eeq}{\end{equation}}
\newcommand{\bieee}{\begin{IEEEeqnarray}{rCl}}
\newcommand{\eieee}{\end{IEEEeqnarray}}
\newcommand{\ii}{\mathrm{i}}

\newcommand{\nblk}{N_\mathrm{blk}}
\newcommand{\nwin}{N_\mathrm{win}}

\newcommand{\citeapostrophe}[1]{\citeauthor{#1}'s \citeyearpar{#1}}

\title{(Why) Jets quantise trapped acoustic modes}

\author{Brandon Yeung\aff{1},
Michael N. Stavropoulos\aff{2},
Vincent Jaunet\aff{2},
Peter Jordan\aff{2},
 \and Oliver T. Schmidt\aff{1}
  \corresp{\email{oschmidt@ucsd.edu}}}

\affiliation{\aff{1}Department of Mechanical and Aerospace Engineering, University of California San Diego, La Jolla, CA 92093, USA
\aff{2}D\'epartement Fluides, Thermique et Combustion, Institut Pprime, CNRS -- Universit\'e de Poitiers --
ISAE-ENSMA, 86000 Poitiers, France}

\begin{document}
\maketitle

\begin{abstract}
High-subsonic turbulent jets trap acoustic modes in their potential core. The modes lie in narrow frequency bands that are $O(0.1)$-wide in Strouhal number, $St$. By studying a Mach 0.9, Reynolds number $10^6$ jet, this work confirms an earlier hypothesis: each band is made up of $O(10)$ discrete frequencies, $\Delta St\sim O(0.01)$ apart. To resolve them, we collect $O(100\,000)$ snapshots from dedicated schlieren experiments and large-eddy simulations (LES), and apply spectral proper orthogonal decomposition (SPOD). Our SPOD algorithm uses sine tapers and enforces the jet's rotational and reflectional symmetries. For both the schlieren and the LES, the trapped SPOD modes are bounded in the $St$--$x$ and $St$--$k$ planes, where $x$ and $k$ are respectively the axial coordinate and wavenumber. The bounds quantise the modes in $St$, $x$, and $k$, such that the $n_x$th mode in $St$ has $n_x$ antinodes along $x$ and $k$.
We then predict trapped modes with local, global, modal, and non-modal stability theories. We find evidence of a competition between modal and non-modal stability.
From only the theoretical bounds, we build a simple in-phase standing-wave resonance model that predicts the quantised global-mode frequencies.

\end{abstract}

\section{Introduction}
Coherent structures in turbulent jets order apparently-disordered motions \citep{Mollo-Christensen1963AGARD,CrowChampagne1971JFM}.
That ability has spurred many experiments and simulations to find such structures, and models to explain them \citep{JordanColonius2013AnnuRev}. The ones we pursue are the trapped acoustic modes \citet{Michalke1970DLR} and \citet{TamHu1989JFM} discovered in round jets. As \citet{TowneEtAl2017JFM} and \citet{SchmidtEtAl2017JFM} showed, at Mach numbers $0.82<M<1$, these modes resonate in the jet core. They belong to discrete azimuthal--radial wavenumber pairs in narrow frequency bands.

From linear theories, \citet{SchmidtEtAl2017JFM} and \citet{TowneEtAl2017JFM,TowneEtAl2019AIAA} hypothesised further that each frequency band is made up of not a continuum of frequencies, but tightly packed discrete values owing to axial resonance. However, their hypothesis remains unverified by data.

In this work, we confirm the discreteness theorised is physical. Our evidence, introduced in \S\S \ref{subsec:EXP}--\ref{subsec:SPOD}, consists of extensive, new schlieren and large-eddy simulation (LES) data from a $M=0.9$ jet \citep{BresEtAl2018JFM}. We apply to the data a highly resolving algorithm \citep{YeungSchmidt2024TCFD} for spectral proper orthogonal decomposition \citep[SPOD;][]{Lumley1970AP,TowneEtAl2018JFM,SchmidtColonius2020AIAAJ}. Section \ref{subsec:theories} gathers linear models: parallel, non-parallel, modal, and non-modal. They explain that trapped modes are quantised by standing-wave resonance \citep{JordanEtAl2018JFM} between axial end conditions. In \S \ref{sec:results}, we present the evidence and test the model predictions. Our findings are then discussed and summarised in \S\S \ref{sec:discussion}--\ref{sec:conclusion}.

\section{Methodology}
For the Mach 0.9, Reynolds number $10^6$, isothermal, round, and turbulent jet that has been studied using experiments, LES, and theory \citep[e.g. by][]{BresEtAl2018JFM,TowneEtAl2017JFM,SchmidtEtAl2017JFM,SchmidtEtAl2018JFM}, we conduct a new experimental--numerical--theoretical campaign. This section outlines the methods that carry over from the earlier studies, and describes the novelties more carefully.

\begin{figure}
\centering
\begin{tikzpicture}[font=\scriptsize]
\node (graphics) {\includegraphics{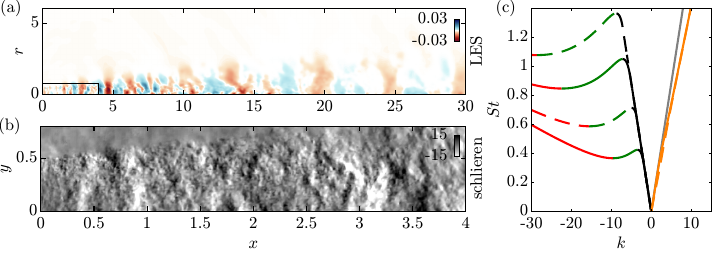}};
\draw[neonCyan,very thick] (-5.35,-1.45) -- ++(7.2,0);
\draw[neonCyan,very thick,dash pattern=on7off5] (-5.35,-.995) -- ++(7.2,0);
\draw[neonCyan,->,>=Latex,semithick] (-5.8,-.995) -- node[left,pos=0]{figures \ref{fig:freq_x_SCHLES}--\ref{fig:freq_kx_SCHLES}} ++(.4,0); \draw[neonCyan,->,>=Latex,semithick] (-5.8,-.995) -- ++(.4,-.37);
\draw[neonCyan,semithick] (-5.35,.53) -- ++(1,0);
\draw[neonCyan,semithick,dash pattern=on7off5] (-5.35,.59) -- ++(1,0);
\node[red] at (3.3,.3) {$k_d^-$};
\node[soberGreen] at (4.2,.28) {$k_T^+$};
\node at (4.9,.28) {$k_p^-$};
\node[neonOrange] at (5.6,-.2) {$k_{\it KH}^+$};
\node[gray] at (5.3,1.8) {$c_j$};
\node[rotate=20] at (3.9,-.9) {$(m,n_r)=(0,1)$};
\node[rotate=20] at (4.3,-.1) {$(1,1)$};
\node[rotate=20] at (4.2,.7) {$(0,2)$};
\node[rotate=20] at (4.1,1.4) {$(1,2)$};
\node at (-4.85,1.8) {$\varOmega_{\it LES}$};
\node at (-4.85,.9) {$\varOmega_{\it sch}$}; \node[white] at (-4.85,-.2) {$\varOmega_{\it sch}$};
\end{tikzpicture}
\caption{Instantaneous fluctuations of $m=0$ component of the LES pressure (a) and symmetric component of the schlieren (b). Their fields of view are $\varOmega_{\it LES}$ and $\varOmega_{\it sch}$, respectively. Teal lines mark $y,r=0$ (solid) and $0.25$ (dashed); see figures \ref{fig:freq_x_SCHLES} and \ref{fig:freq_kx_SCHLES}. Panel (c) shows vortex-sheet dispersion relations made up of neutral, propagative $k_d^-$, $k_T^+$, and $k_p^-$ modes and (the real part of) an unstable $k_{\it KH}^+$ mode, with the sound speed, $c_j$, for reference. Solid and dashed curves respectively denote $m=0$ and 1.}
\label{fig:overview}
\end{figure}
\begin{table}
\begin{center}
\def~{\hphantom{0}}
\begin{tabular}{lcccccccc}
Database & $N_x\times N_{y,r}(\times\ N_\theta)$ & $\Delta t c_j/D$ & $N_t$ & $N_f$ & $N_\mathrm{ovlp}$ & $N_\mathrm{blk}$ & $N_\mathrm{win}$ & $N_\mathrm{ens}$ \\[3pt]
Schlieren & 464\,$\times$\,186 & 0.345 & 101\,041 & 3368 & 0 & 30 & 6 & 180 \\
LES & $656\times138\times128$ & 0.2 & 240\,000 & 4000 & 0 & 60 & 4 & 240
\end{tabular}
\caption{Database and spectral estimation parameters. Throughout this work, lengths are non-dimensionalised by the diameter, $D$. Frequencies are reported as Strouhal numbers, $St=fD/U_j=\omega/(2\pi M)$, where $\omega$ is the angular frequency, $M=U_j/c_j=0.9$ is the Mach number, and $U_j$ and $c_j$ are respectively the jet exit and jet sound speeds.}
\label{tab:params}
\end{center}
\end{table}

\subsection{Schlieren experiments}\label{subsec:EXP}
We perform schlieren experiments at the \textit{Bruit et Vent} jet-noise facility of the Institut Pprime in Poitiers, France. The jet flow issues from a convergent--straight nozzle with a diameter of $D=0.05\ \mathrm{m}$ and a tripped internal boundary layer. The flow is imaged with a Phantom v2640 camera at a sampling frequency of 20 kHz for 5.05 s. This creates $N_t=101\,041$ images with $1024\times976$ pixels each. The images are cropped to $x/D\in[0,4]$ and $y/D\in[-0.8,0.8]$ in the horizontal and vertical directions, respectively, smoothed to reduce noise, and downsampled by a factor of two in each spatial direction, resulting in a resolution of $N_x \times N_y = 464\times186$. Table \ref{tab:params} summarises these parameters, along with the non-dimensionalisation we use throughout. Each image is further decomposed into components symmetric and antisymmetric about $y=0$. Because schlieren is line-of-sight integrated, each component is a mixture of (integer) azimuthal wavenumbers, $m$. Figure \ref{fig:overview}(b) displays a snapshot of the symmetric component.

\subsection{Large-eddy simulations}\label{subsec:LES}
We carry out LES using the unstructured, compressible solver `Charles' developed by Cadence \citep{BresEtAl2017AIAAJ}. Our set-up is identical to \citeapostrophe{BresEtAl2018JFM} LES, which was validated against the experiments at Pprime. The computational domain includes the nozzle. Its internal boundary layer is tripped by synthetic turbulence. To reduce wall-clock time, we run three simulations in parallel. The density, $\rho$, temperature, $T$, and velocities, $u_x$, $u_r$, and $u_\theta$, non-dimensionalised by the jet exit conditions, are saved at a time interval of $\Delta t c_j/D=0.2$ to a total of $N_t=240\,000$ snapshots. The snapshots are interpolated onto an $N_x \times N_r \times N_\theta=656\times138\times128$ cylindrical grid that spans $x/D\in[0,30]$, $r/D\in[0,6]$, and $\theta\in[0,2\pi)$ in the axial, radial, and azimuthal directions, respectively. Statistical axisymmetry means azimuthal-wavenumber components are uncorrelated and therefore can be analysed separately \citep[e.g.]{Lumley1970AP}. Figure \ref{fig:overview}(a) shows a snapshot of the $m=0$ pressure. Throughout this work (except for the schlieren), we will always visualise the pressure field.

\subsection{Spectral proper orthogonal decomposition}\label{subsec:SPOD}
At its heart, SPOD \citep{Lumley1970AP,TowneEtAl2018JFM,SchmidtColonius2020AIAAJ} solves the eigenvalue problem (EVP)
\begin{equation}\label{eq:SPODcontinuous}
\int_\varOmega \mathsfbi S(\vb*x,\vb*x',f) \mathsfbi W(\vb*x') \vb*{\phi}_j(\vb*x',f) \dd{\vb*x'} = \lambda_j(f) \vb*{\phi}_j(\vb*x,f)
\end{equation}
of the cross-spectral density (CSD) tensor, $\mathsfbi S = E\{ \hat{\vb*q}(\vb*x,f) \hat{\vb*q}^*(\vb*x',f) \}$, at frequency, $f$, for the eigenvalue and mode, $\lambda_j$ and $\vb*{\phi}_j$, respectively. The operator $E\{\cdot\}$ denotes the expectation, $\hat{(\cdot)}$ the Fourier transform, and $(\cdot)^*$ the conjugate transpose; $\varOmega$ is a domain of interest and $\mathsfbi W$ a weight. The eigenvalue measures energy in the norm, $\|\cdot\|_{\vb*x}$, induced by the inner product $\langle\vb*q_1,\vb*q_2\rangle_{\vb*x}=\int_\varOmega \vb*q_2^* \mathsfbi W \vb*q_1 \dd{\vb*x}$. For the schlieren data, the univariate state, $q$, contains image pixels and $\mathsfbi W$ is superfluous. For the LES data, the state is $\vb*q=[\rho,T,u_x,u_r,u_\theta]^\mathrm{T}$, where $(\cdot)^\mathrm{T}$ denotes the transpose, and
\begin{equation}\label{eq:chuWeightTensor}
\mathsfbi W = \mathrm{diag}([\bar T/(\gamma\bar\rho M^2),\bar\rho/(\gamma(\gamma-1)\bar TM^2),\bar\rho,\bar\rho,\bar\rho])
\end{equation}
leads to the compressible-energy norm \citep{Chu1965ActaMech}. Here, $\bar{(\cdot)}$ denotes the mean and $\gamma$ the adiabatic constant. When we solve \eqref{eq:SPODcontinuous}, the physics of trapped modes prompts two subtleties. First, trapped modes are spatially compact \citep{TowneEtAl2017JFM,SchmidtEtAl2017JFM}. To ensure they are a meaningful portion of the total energy in the induced norm, we restrict $\varOmega$ to $x,y\in[0,3]\times[0,0.3]$ and $x,r\in[0,4]\times[0,0.25]$ for the schlieren and LES, respectively. Only the energy within $\varOmega$ is measured by the eigenvalues. 
But the modes extend to the whole domain, $\varOmega_{\it LES}$ or $\varOmega_{\it sch}$, in figure \ref{fig:overview}(a,b) \citep{Boree2003ExpFluids}. Second, the frequency separation between the trapped global modes \citet{SchmidtEtAl2017JFM} identified is minute, as small as $\Delta St \sim O(0.01)$. To distinguish between these waves in data, we gather long databases and apply sine-taper SPOD \citep{YeungSchmidt2024TCFD}, which estimates $\hat{\vb*q}$ from sinusoidal window functions \citep{RiedelSidorenko1995IEEE}. Our estimation parameters, i.e., the block length, $N_f$, overlap between blocks, $N_\mathrm{ovlp}$, number of blocks, $N_\mathrm{blk}$, number of windows, $N_\mathrm{win}$, and ensemble size, $N_\mathrm{ens} = N_\mathrm{blk}N_\mathrm{win}$, are reported in table \ref{tab:params}.

As \S\S \ref{subsec:EXP}--\ref{subsec:LES} stated, we decompose snapshots into reflectional (schlieren) or azimuthal (LES) symmetry components and treat each component separately. The three-dimensional LES dataset has another statistical symmetry: invariance with respect to reflection in $\theta$ or, equivalently, inversion of $m$ \citep{Sirovich1987QAM2,MoinMoser1989JFM,pope2000cambridge,JungEtAl2004JFM,SmithEtAl2005,HolmesEtAl2012Cambridge,DavoustEtAl2012JFM,SchmidtEtAl2018JFM,MagriEtAl2023AnnuRev}. We define the transformation
\begin{subequations}
\begin{align}
\mathcal{R}_m([\hat\rho,\hat T,\hat u_x,\hat u_r,\hat u_\theta]^\mathrm{T}(\vb*x,m,f)) &= [\hat\rho,\hat T,\hat u_x,\hat u_r,-\hat u_\theta]^\mathrm{T}(\vb*x,-m,f) \label{subeq:2dDFTSym_a}\\
&= [\hat\rho,\hat T,\hat u_x,\hat u_r,-\hat u_\theta]^*(\vb*x,m,-f), \label{subeq:2dDFTSym_b}
\end{align}
\end{subequations}
where $\hat{(\cdot)}$ is now the azimuthal--temporal Fourier transform and $\vb*x$ represents $(x,r)$. \eqref{subeq:2dDFTSym_b} follows from the conjugate symmetry of real $\vb*q$. On the CSD, $\mathsfbi S_m = E\{ \hat{\vb*q}(\vb*x,m,f) \hat{\vb*q}^*(\vb*x',m,f) \}$, $\mathcal{R}_m$-invariance imposes the symmetry
\begin{subequations}\label{eq:CSDm-nonzero}
\begin{align}
E\{ \hat{\vb*q}(\vb*x,m,f) \hat{\vb*q}^*(\vb*x',m,f) \} &= E\{ \mathcal{R}_m( \hat{\vb*q}(\vb*x,m,f) ) \mathcal{R}_m( \hat{\vb*q}(\vb*x',m,f) )^* \}, \\
E\qty{ \mqty[\hat{\vb*q}_{1:4}\hat{\vb*q}_{1:4}^* & \hat{\vb*q}_{1:4}\hat u_\theta^* \\
\hat u_\theta\hat{\vb*q}_{1:4}^* & \hat u_\theta\hat u_\theta^* ] }(\vb*x,\vb*x',m,f)
&= E\qty{ \mqty[\hat{\vb*q}_{1:4}\hat{\vb*q}_{1:4}^* & -\hat{\vb*q}_{1:4}\hat u_\theta^* \\
-\hat u_\theta\hat{\vb*q}_{1:4}^* & \hat u_\theta\hat u_\theta^* ] }(\vb*x,\vb*x',-m,f). \IEEEeqnarraynumspace \label{subeq:CSDm-nonzero}
\end{align}
\end{subequations}
\eqref{subeq:CSDm-nonzero} splits $\hat{\vb*q}$ into $\hat{\vb*q}_{1:4}=[\hat\rho,\hat T,\hat u_x,\hat u_r]^\mathrm{T}$ and $\hat u_\theta$. In practice, enforcing \eqref{eq:CSDm-nonzero} requires nothing more than building the data matrix
\begin{equation}
\hat{\mathsfbi Q} = [\hat{\vb*q}^{(1)},\ldots,\hat{\vb*q}^{(N_\mathrm{ens})}, \underbrace{\mathcal{R}_m(\hat{\vb*q}^{(1)}),\ldots,\mathcal{R}_m(\hat{\vb*q}^{(N_\mathrm{ens})})}_\text{\eqref{subeq:2dDFTSym_a} or, more conveniently, \eqref{subeq:2dDFTSym_b}} ],
\end{equation}
from which it is clear the ensemble size has doubled.

For $m=0$, \eqref{subeq:CSDm-nonzero} immediately simplifies $\mathsfbi S_m$ to the block-diagonal structure
\begin{equation}
\mathsfbi S_0 = \mqty[ E\{ \hat{\vb*q}_{1:4} \hat{\vb*q}_{1:4}^* \} & \vb 0 \\ \vb 0^\mathrm{T} & E\{ \hat u_\theta\hat u_\theta^* \} ](\vb*x,\vb*x',0,f),
\end{equation}
which reduces to two decoupled, smaller EVPs respectively of $E\{ \hat{\vb*q}_{1:4} \hat{\vb*q}_{1:4}^* \}$ and $E\{ \hat u_\theta\hat u_\theta^* \}$.

\subsection{Linear theories}\label{subsec:theories}
We will compare the trapped modes from SPOD and the predictions of the following linear models.

\subsubsection{Local stability}\label{subsubsec:VS}
The simplest model we consider, derived from the locally parallel, linearised Euler equations, is the cylindrical vortex sheet. It has the dispersion relation \citep{LessenEtAl1965JFM,Michalke1970DLR,TamHu1989JFM,TowneEtAl2017JFM}
\begin{equation}\label{eq:VSdispersion}
\mathcal{D}(k,\omega; M, T) = \frac{1}{\qty(1-\frac{kM}{\omega})^2} + \frac{1}{T}\frac{\mathrm I_m(\frac{\gamma_i}{2})\qty[\frac{\gamma_o}{2}\mathrm K_{m-1}(\frac{\gamma_o}{2})+m\mathrm K_m(\frac{\gamma_o}{2})]}{\mathrm K_m(\frac{\gamma_o}{2})\qty[\frac{\gamma_i}{2}\mathrm I_{m-1}(\frac{\gamma_i}{2})-m\mathrm I_m(\frac{\gamma_i}{2})]} = 0,
\end{equation}
where $\mathrm I_m$ and $\mathrm K_m$ are $m$th-order modified Bessel functions respectively of the first and second kind, $\gamma_i=\sqrt{k^2-(\omega-Mk)^2/T}$, $\gamma_o=\sqrt{k^2-\omega^2}$, and $k$ is the axial wavenumber. The eigenvalues of \eqref{eq:VSdispersion} belong to several families of waves, summarised in figure \ref{fig:overview}(c), where we have adopted \citeapostrophe{JordanEtAl2018JFM} nomenclature. In the $k>0$ half-plane resides the well-known spatially-unstable Kelvin--Helmholtz (KH) mode, denoted $k^+_{\it KH}$. In the $k<0$ half-plane are the trapped modes we seek. They are neutrally stable, propagative modes given by real $\omega$ and $k$ and form curved branches, each linked with a discrete azimuthal--radial wavenumber pair \citep{TowneEtAl2017JFM,SchmidtEtAl2017JFM}, $(m,n_r)$. We will focus on the first two, i.e., $(m,n_r)=(0,1)$ and $(1,1)$. Each branch consists of three modes: an upstream- and a downstream-propagating duct-like mode, $k^-_d$ and $k^+_T$, respectively, and an upstream-propagating acoustic jet mode, $k^-_p$. No further discreteness, beyond that in $(m,n_r)$, is permitted by parallel theory because, as we shall see, axial quantisation is contingent on non-parallelism. For that, we need the (weakly) non-parallel theories introduced next. 

\subsubsection{Weakly non-parallel stability and standing waves}\label{subsubsec:frankensteinTheory}
\begin{figure}
\centering
\begin{tikzpicture}[font=\scriptsize]
\node (graphics) {\includegraphics{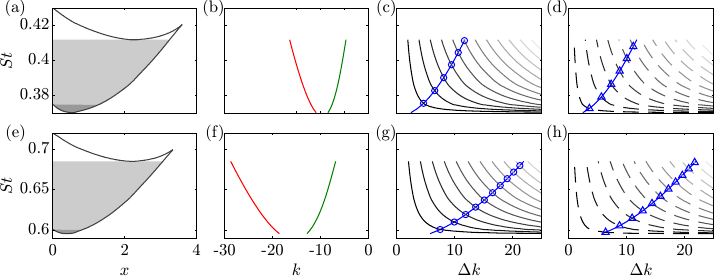}};
\node[rotate=90] at (6.25,1.25) {$m=0$}; \node[rotate=90] at (6.25,-.75) {$m=1$};
\draw[<->,>=Latex] (-5.2,1) -- node[above]{$L(\omega)$} ++(1.45,0);
\draw[<->,>=Latex] (-5.2,-1.2) -- node[above]{$L(\omega)$} ++(1.1,0);
\node at (-3.3,1.2) {S1}; \node at (-4.5,2) {S2};
\node at (-3.5,-.9) {S1}; \node at (-4.5,-.1) {S2};
\node[Red] at (-1.25,1) {$k_d^-$}; \node[soberGreen] at (-.1,1) {$k_T^+$};
\node[Red] at (-2,-1.2) {$k_d^-$}; \node[soberGreen] at (-.4,-1.2) {$k_T^+$};
\draw[<->,>=Latex,blue] (-1,1) -- node[above]{$\Delta k(\omega)$} ++(.7,0);
\draw[<->,>=Latex,blue] (-1.7,-1.2) -- node[above]{$\Delta k(\omega)$} ++(1.1,0);
\node at (1.1,1.9) {$\dfrac{2n_x\pi}{L(\omega)}$}; \node[blue] at (2,1.9) {$\Delta k(\omega)$};
\node at (1.1,-.2) {$\dfrac{2n_x\pi}{L(\omega)}$}; \node[blue] at (2.7,-.2) {$\Delta k(\omega)$};
\node at (4.2,1.9) {$\dfrac{(2n_x+1)\pi}{L(\omega)}$}; \node[blue] at (5.4,1.9) {$\Delta k(\omega)$};
\node at (4.2,-.2) {$\dfrac{(2n_x+1)\pi}{L(\omega)}$}; \node[blue] at (5.6,-.2) {$\Delta k(\omega)$};
\draw[->,>=Latex] (2,.5) -- node[sloped,pos=.9,right]{$n_x$} ++(.7,.7);
\draw[->,>=Latex] (2,-1.65) -- node[sloped,pos=.9,right]{$n_x$} ++(.7,.7);
\draw[->,>=Latex,dashed] (4.9,.5) -- node[sloped,pos=.9,right]{$n_x$} ++(.7,.7);
\draw[->,>=Latex,dashed] (4.9,-1.65) -- node[sloped,pos=.9,right]{$n_x$} ++(.7,.7);
\end{tikzpicture}
\caption{Standing-wave resonance: (a--d) $m=0$; (e--h) $m=1$. The saddle points, S1 and S2, in (a,e) were computed by \citet{TowneEtAl2017JFM} from a weakly non-parallel, linearised Euler model. Propagative $k_d^-$ and $k_T^+$ waves are reflected between $x=0$ and an S1 saddle (light grey) or between two S1 saddles (dark grey). The vortex-sheet eigenvalues in (b,f) come from figure \ref{fig:overview}(c). In (c,d,g,h), lighter grey lines represent larger $n_x$. In-phase (c,g) and out-of-phase (d,h) resonances respectively admit the standing waves marked by circles and triangles.}
\label{fig:standingWavePredictions}
\end{figure}
Relaxing the parallel-flow assumption in \S \ref{subsubsec:VS}, WKBJ theory supplies one explanation of axial quantisation. A step in this direction was taken by \citet{TowneEtAl2017JFM} and \citet{SchmidtEtAl2017JFM}, who found turning points connecting upstream- and downstream-propagating waves at the nozzle plane and two saddle points. The saddles they calculated, labelled S1 and S2, are reproduced in figure \ref{fig:standingWavePredictions}(a,e) for the $(m,n_r)=(0,1)$ and $(1,1)$ branches. By matching the WKBJ approximations for two turning points, e.g. $x=0$ and S1, it is possible to derive a quantisation condition for a discrete set of global approximations \citep{BenderOrszag1999Springer}.

Instead of this rigorous but elaborate approach, we seek a resonance model that is simpler yet predictive. Our model rests on the notion \citep[and others]{JordanEtAl2018JFM,NogueiraEtAl2024JFM}
that standing waves, confined to some length scale, $L$, choose discrete frequencies. We express it as the conditions
\begin{subequations}\label{eq:standingWave}
\begin{align}
\Delta k(\omega) &= \frac{2n_x\pi}{L(\omega)} \qc n_x=1,2,\ldots \qc \text{for in-phase resonance}, \label{subeq:standingWaveInPhase}\\ \qor
\Delta k(\omega) &= \frac{(2n_x+1)\pi}{L(\omega)} \qc n_x=0,1,\ldots \qc \text{for out-of-phase resonance}, \label{subeq:standingWaveOutPhase}
\end{align}
\end{subequations}
where $\Delta k(\omega)=k_T^+(\omega)-k_d^-(\omega)$ comes from the vortex-sheet dispersion relation \eqref{eq:VSdispersion} and $n_x$ is an axial wavenumber index. Here, $L(\omega)$ is the distance between turning points and therefore varies with frequency; see figure \ref{fig:standingWavePredictions}(a,e). We consider the simplest type of resonance, enabled by reflections of $k_d^-$ and $k_T^+$ waves at $x=0$ and an S1 saddle or, for low frequencies, at two S1 saddles. The region where it may occur is shaded in grey. Above the shaded region, several resonances are possible \citep{TowneEtAl2017JFM}, creating ambiguity in the waves and turning points at work, so our model excludes them. Finally, solving \eqref{eq:standingWave} for $\omega$ yields the predictions in panels \ref{fig:standingWavePredictions}(c,g) and (d,h) for in- and out-of-phase resonances, respectively. In- and out-of-phase resonances each predict six standing waves for $m=0$; they predict, respectively, 10 and 11 waves for $m=1$. In every solution except for $n_x=0$, the upstream turning point is $x=0$, not S1. Intuitively, the in-phase (out-of-phase) waves are akin to resonant pressure modes in a finite but extensible tube with both ends (one end) open \citep[e.g.]{Crawford1968Berkeley}.

\subsubsection{Global modal and non-modal stability}\label{subsubsec:globalTheory}
To account fully for non-parallelism, we next turn to global stability \citep{NicholsLele2011JFM,SchmidtEtAl2017JFM}, modal and non-modal.
For a discretised compressible Navier--Stokes operator, $\mathsfbi A_m$, linearised about the turbulent mean \citep{CrightonGaster1976JFM,McKeonSharma2010JFM,JordanColonius2013AnnuRev}, modal analysis solves the EVP
\begin{equation}\label{eq:globalModalStab}
(\ii\omega\mathsfbi I + \mathsfbi A_m) \hat{\vb*q} = 0
\end{equation}
for modes, $\hat{\vb*q}$, and eigenvalues, $\omega\in\mathbb C$. Each eigenvalue encodes in its real and imaginary parts an angular frequency, $\omega_r=2\pi M St$, and growth rate, $\omega_i$, respectively. Whereas a local eigenmode (\S\S \ref{subsubsec:VS}--\ref{subsubsec:frankensteinTheory}) has but one wavenumber, a global mode, like an SPOD mode, may have a spectrum of wavenumbers, e.g. those taking part in resonance.

For the same $\mathsfbi A_m$, non-modal analysis finds the level curves of the norm of the resolvent, $(z\mathsfbi I - \mathsfbi A_m)^{-1}$, i.e., the $\varepsilon$-pseudospectrum \citep{ReddyEtAl1993SIAM,TrefethenEtAl1993Science,SchmidHenningson2001Springer,TrefethenEmbree2005PUP}. Given the compressible-energy norm, $\|\cdot\|_{\vb*x}$, induced from \eqref{eq:chuWeightTensor}, we solve
\begin{equation}\label{eq:globalNonmodalStab}
\|(z\mathsfbi I - \mathsfbi A_m)^{-1}\|_{\vb*x} > \varepsilon^{-1} \qq{with} \varepsilon>0.
\end{equation}
The $\varepsilon$-pseudoeigenvalues, $z\in\mathbb C$, are equivalent to the eigenvalues of a randomly perturbed operator, $\mathsfbi A_m + \mathsfbi E$, with $\|\mathsfbi E\|_{\vb*x}<\varepsilon$. A highly non-normal $\mathsfbi A_m$ permits a large resolvent norm even when $z$ is distant from its eigenvalues; equivalently, its eigenvalues may be sensitive to perturbation and, if viewed in isolation, deceptive.

To build $\mathsfbi A_m$ and compute \eqref{eq:globalModalStab} and \eqref{eq:globalNonmodalStab}, we use the linear solver `Linstab2D' \citep{SchmidtRist2011JFM,MartiniSchmidt2024TCFD}. The operating conditions, including the Reynolds number, match the experiment's and LES's. The domain is discretised along the axial and radial directions by $2300\times250$ points, respectively. It excludes the nozzle. Instead, an inlet sponge mimics the nozzle impedance, a tactic \citet{SchmidtEtAl2017JFM} justified. Moreover, they showed trapped global modes to be largely indifferent to variations in turbulence modelling. We thus choose a basic eddy-viscosity model. Denoting the Frobenius inner product by $\langle\cdot,\cdot\rangle_F$, we define the eddy-viscosity field $\bar\nu_t(\vb*x) = \langle-\bar{\mathsfbi D}(\vb*x),\bar{\mathsfbi S}(\vb*x)\rangle_F / \langle\bar{\mathsfbi S}(\vb*x),\bar{\mathsfbi S}(\vb*x)\rangle_F$, which needs only spatially-resolved mean quantities: the deviatoric Reynolds-stress tensor, $\bar{\mathsfbi D}$, and strain-rate tensor, $\bar{\mathsfbi S}$ \citep[for details, see e.g.][]{TammisolaJuniper2016JFM,vonSaldernEtAl2023AIAA}. We compute the eigenvalues of \eqref{eq:globalModalStab} in the range $0.3\lesssim St\lesssim1.4$. Knowing the eigenvalues, we calculate the resolvent norm in \eqref{eq:globalNonmodalStab} in the neighbourhood of the $(m,n_r)=(0,1)$ branch, $0.36\lesssim St\lesssim0.43$ and $-0.013\lesssim \omega_i/(2\pi M)\lesssim0.003$, and $(m,n_r)=(1,1)$ branch, $0.57\lesssim St\lesssim0.72$ and $-0.029\lesssim \omega_i/(2\pi M)\lesssim0.006$. Otherwise our set-up follows \citet{SchmidtEtAl2017JFM}.

\section{Results}\label{sec:results}
In this section we answer three questions about the resonating trapped modes. First, along each $(m,n_r)$ branch, are they discrete (\S \ref{subsec:dataIsQuantised})? Second, if so, why (\S\S \ref{subsec:whyQuantised}--\ref{subsec:modalNonmodalContest})? Third, how well are they predicted by linear theories (\S \ref{subsec:modelsHowGood})? We begin with the most fundamental.
\subsection{Discreteness is physical}\label{subsec:dataIsQuantised}
\begin{figure}
\centering
\begin{tikzpicture}[font=\footnotesize]
\node (graphics) {\includegraphics{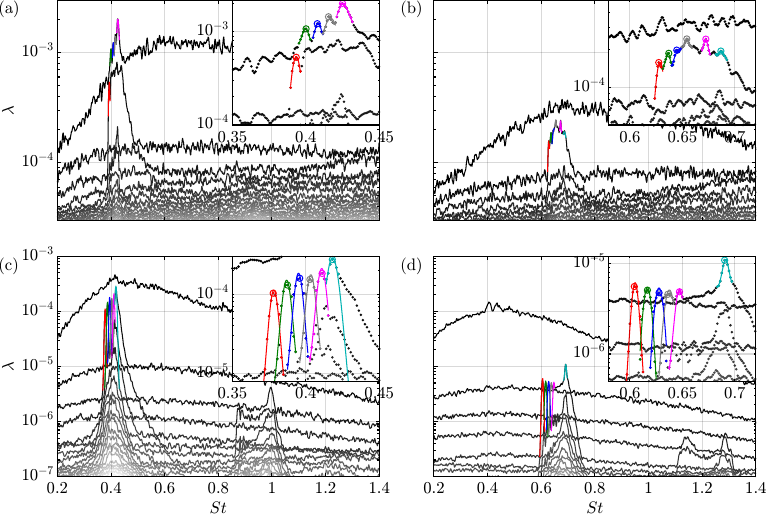}};
\node at (-2.75,4.55) {symmetric}; \node at (3.6,4.55) {antisymmetric};
\node[rotate=90] at (6.6,2.5) {schlieren};
\node at (-2.75,.25) {$m=0$}; \node at (3.6,.25) {$m=1$};
\node[rotate=90] at (6.6,-1.7) {LES};
\end{tikzpicture}
\caption{SPOD spectra of the schlieren's symmetric (a) and antisymmetric (b) components and the LES's $m=0$ (c) and $m=1$ (d) components. In the insets, eigenvalues are drawn as dots and peaks are marked by circles. Modes similar to that at each peak are delineated by colour and fitted with a Gaussian curve.}
\label{fig:cmp_spectra}
\end{figure}
Figure \ref{fig:cmp_spectra} compares the SPOD eigenvalues of the schlieren and LES data. With the exception of figure \ref{fig:cmp_spectra}(a), a broadband spectrum representing the shear-layer KH instability \citep{SchmidtEtAl2018JFM} dominates over all frequencies. The tones in the ranges $0.35\lesssim St\lesssim 0.45$ and $0.6\lesssim St\lesssim 0.7$ in figures \ref{fig:cmp_spectra}(a,c) and \ref{fig:cmp_spectra}(b,d), respectively, are trapped modes along the $(m,n_r)=(0,1)$ and $(1,1)$ branches \citep{SchmidtEtAl2017JFM,TowneEtAl2017JFM}. Although the schlieren measurement blends azimuthal wavenumbers, the symmetric--antisymmetric decomposition separates the two branches. Relative to the LES, in the schlieren the trapped-mode bands are shifted to slightly higher frequencies, which \S \ref{sec:discussion} will discuss. Only the LES shows the $n_r=2$ branches, in the ranges $0.9\lesssim St\lesssim 1$ and $1.1\lesssim St\lesssim 1.3$ for $m=0$ and $1$, respectively.

Figure \ref{fig:cmp_spectra}'s insets, which focus on the $n_r=1$ branches, establish our main hypothesis: trapped modes are discrete.
Their frequency spacing is a mere $\Delta St\sim O(0.01)$. While SPOD eigenvalues of equal rank are typically interpreted as a single flow mechanism and, accordingly, joined in a curve, trapped waves are a counterexample where one mechanism may span several ranks. We demonstrate this by calculating the similarity, $|\langle\vb*\phi_\mathrm{peak},\vb*\phi\rangle_{\vb*x}| \in [0,1]$, between the SPOD mode, $\vb*\phi_\mathrm{peak}$, at each peak (circles in the insets) and SPOD modes at all frequencies and ranks, $\vb*\phi$. The inner product, $\langle\cdot,\cdot\rangle_{\vb*x}$, is the same as the SPOD's. Modes that meet the threshold $|\langle\vb*\phi_\mathrm{peak},\vb*\phi\rangle_{\vb*x}| \ge 0.7$ form clusters in the spectra.
Estimated from data, each cluster inevitably spreads over a few frequency bins, rather than just one. To highlight the clusters, we fit them with Gaussian curves. For the schlieren, perhaps owing to limited statistical convergence,
clusters are confined to either rank one or rank two. For the LES, sharply defined clusters peak in rank two but persist up to rank five. Appendix \ref{sec:appConfidence} calculates confidence intervals around the eigenvalue estimates, showing excellent convergence of the LES SPOD and good convergence of the symmetric schlieren SPOD.

\subsection{Axial boundaries admit discrete axial states}\label{subsec:whyQuantised}
\begin{figure}
\centering
\begin{tikzpicture}[font=\scriptsize]
\node (graphics) {\includegraphics{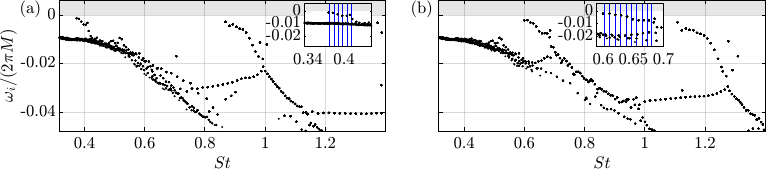}};
\node[rotate=90] at (.25,.4) {$m=0$}; \node[rotate=90] at (6.7,.4) {$m=1$};
\end{tikzpicture}
\caption{Global stability spectra: (a) $m=0$; (b) $m=1$. In the insets, blue lines mark the in-phase standing-wave frequencies from \eqref{subeq:standingWaveInPhase}.}
\label{fig:cmp_linear_theories}
\end{figure}
After confirming that trapped waves are discrete, we now explain how they become so. By comparing standing-wave resonance (\S \ref{subsubsec:frankensteinTheory}) and global modal stability (\S \ref{subsubsec:globalTheory}),
we aim to show that standing waves are a distillation of the non-parallel effects producing discreteness.

In figure \ref{fig:cmp_linear_theories}, as \citet{SchmidtEtAl2017JFM} also found, all global eigenvalues are damped. Closest to the real axis, discrete trapped modes 
form mostly downward-sloping, or increasingly-damped, branches. These modes' confinement to isolated points in the spectrum can be understood from their spatial confinement along two dimensions \citep{SchmidHenningson2001Springer,TrefethenEmbree2005PUP}, $r$ and $x$. That said, because $\mathsfbi A_m$ is discretised, it is hard to infer the point and continuous spectra of the underlying infinite-dimensional operator.

For $m=0$ in figure \ref{fig:cmp_linear_theories}(a)'s inset, the six trapped global modes match the six in-phase standing waves from \eqref{subeq:standingWaveInPhase} in frequency. Similarly, for $m=1$ in figure \ref{fig:cmp_linear_theories}(b)'s inset, the first five (from the left) trapped global modes and in-phase standing waves match, beyond which the approximation gradually worsens. We dismiss the out-of-phase standing-wave frequencies from \eqref{subeq:standingWaveOutPhase} because they are too low, and conclude that in-phase resonance is more representative.
Overall, standing waves convincingly emulate global theory. This supports the claim in \S \ref{subsubsec:frankensteinTheory}: axial boundary conditions, which the nozzle plane and S1 saddle supply, and which only discrete values of $k_T^+$ and $k_d^-$ satisfy, are the main ingredient in quantisation.

\begin{figure}
\centering
\begin{tikzpicture}[font=\scriptsize]
\node (graphics) {\includegraphics{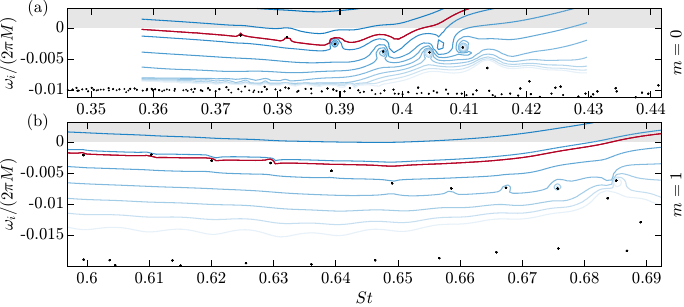}};
\node at (-1.8,1.8) {$n_x=1$};
\node at (-4.2,-.15) {$n_x=1$};
\draw[->,>=Latex,sky] (-3.5,1.4) -- node[left]{$\varepsilon$} ++(0,.5);
\draw[->,>=Latex,sky] (-3.5,-1) -- node[left]{$\varepsilon$} ++(0,.5);
\node[maroon] at (2.5,2.3) {$\varepsilon=10^{-3.96}$};
\node[maroon] at (2.5,-.25) {$\varepsilon=10^{-4.22}$};
\end{tikzpicture}
\caption{Global spectra and $\varepsilon$-pseudospectra: (a) $m=0$; (b) $m=1$. Level curves from light to dark blue show $\varepsilon=10^{-4.7},10^{-4.6},\ldots,10^{-3.8}$ (a) and $10^{-4.7},10^{-4.6},\ldots,10^{-4.1}$ (b). Maroon highlights $\varepsilon=10^{-3.96}$ (a) and $10^{-4.22}$ (b).}
\label{fig:pseudospectra}
\end{figure}

\subsection{Modal and non-modal stability competes}\label{subsec:modalNonmodalContest}
The modal theories in figures \ref{fig:standingWavePredictions} and \ref{fig:cmp_linear_theories} explain why trapped waves are discrete, but not the trends in SPOD energy along a branch. For $(m,n_r)=(0,1)$ in figure \ref{fig:cmp_spectra}(a,c), the peak SPOD eigenvalues nearly all grow with frequency. For $(m,n_r)=(1,1)$ in figure \ref{fig:cmp_spectra}(d), the LES SPOD has sharper peaks (than does the schlieren SPOD), which first decay then grow and are most energetic at high frequency. These trends are at odds with the global spectra in figure \ref{fig:cmp_linear_theories}, where the lower frequencies in each branch are generally less damped. 

One explanation is non-normality. Figure \ref{fig:pseudospectra} overlays the trapped-mode branches with logarithmically-spaced $\varepsilon$-pseudospectra, and shows that higher-frequency trapped modes are more non-normal. For concreteness, we highlight two level curves, $\varepsilon=10^{-3.96}$ and $10^{-4.22}$, respectively for $m=0$ and 1, each passing just above the $n_x=1$ eigenvalue. At these levels of $\varepsilon$, below both $St\approx 0.38$ ($n_x=2$) for $m=0$ and $St\approx 0.63$ ($n_x=4$) for $m=1$, random perturbations negligibly affect stability. But above these frequencies, the level curves reach higher (less negative) $\omega_i$ than the eigenvalues do. That is, non-normality lessens damping.
For $m=0$ and $St\gtrsim 0.403$ and for $m=1$ and $St\gtrsim 0.684$, at the highlighted $\varepsilon$, the pseudoeigenvalues may even be unstable. The least-damped eigenvalues are the furthest from the branch intersections. The most-damped are the closest and, as expected, also the most non-normal \citep{ReddyEtAl1993SIAM}. This leads higher-frequency trapped modes to be modally more damped, but non-modally less so. The trends in SPOD energy suggest the two causes are competing.
For $(m,n_r)=(0,1)$, the energy's rise hints at the non-modal cause dominating. For $(m,n_r)=(1,1)$, the energy falls then rises, which points to both causes, with the non-modal gradually outcompeting the modal.

\subsection{Linear theories are predictive}\label{subsec:modelsHowGood}
\begin{figure}
\centering
\begin{tikzpicture}[font=\scriptsize]
\node (graphics) {\includegraphics{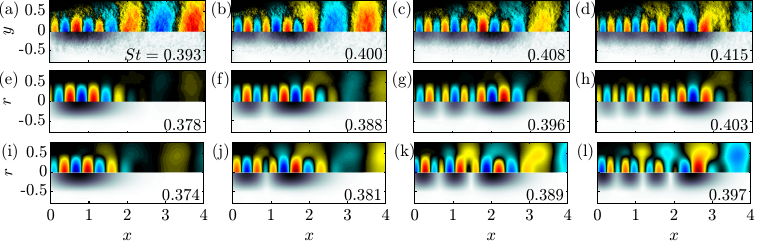}};
\node[rotate=90] at (6.6,1.5) {schlieren};
\node[rotate=90] at (6.6,.4) {LES};
\node[rotate=90,text width=30,align=center] at (6.6,-.9) {global stability};
\end{tikzpicture}
\caption{First four trapped modes, $n_x=1,2,3,4$, of the symmetric schlieren SPOD (a--d), $m=0$ LES SPOD (e--h), and $m=0$ global stability (i--l). The real part of each mode is shown in $y,r\ge0$. The absolute value of the $k\le0$ part is shown in $y,r<0$.}
\label{fig:cmp_modes_m0}
\end{figure}
\begin{figure}
\centering
\begin{tikzpicture}[font=\scriptsize]
\node (graphics) {\includegraphics{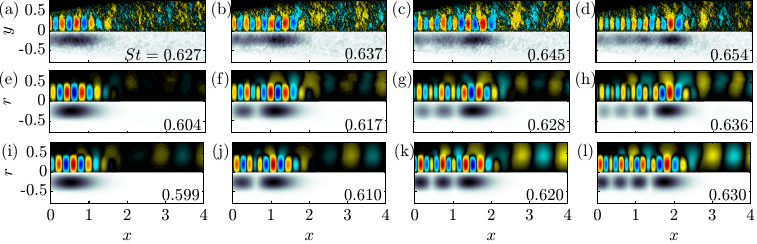}};
\node[rotate=90] at (6.6,1.5) {schlieren};
\node[rotate=90] at (6.6,.4) {LES};
\node[rotate=90,text width=30,align=center] at (6.6,-.9) {global stability};
\end{tikzpicture}
\caption{Same as figure \ref{fig:cmp_modes_m0} but for the antisymmetric and $m=1$ modes.}
\label{fig:cmp_modes_m1}
\end{figure}
So far we have seen that linear theories are qualitative models of trapped-mode physics. Now we assess if they can make quantitative predictions too, specifically of mode frequencies and structures. For both $(m,n_r)$ branches and $n_x=1,2,3,4$, figures \ref{fig:cmp_modes_m0} and \ref{fig:cmp_modes_m1} compare the corresponding global (first four eigenvalues in each inset of figure \ref{fig:cmp_linear_theories}) and SPOD (first four circles in each inset of figure \ref{fig:cmp_spectra}) modes. They mostly match.
The global frequencies underpredict the schlieren and LES frequencies by $\Delta St\lesssim 0.03$ and $0.008$, respectively. As \citet{SchmidtEtAl2017JFM} and \citet{TowneEtAl2017JFM} found, the modulus of each mode, with its $k>0$ component filtered out (keeping $k\le0$) for clarity, has $n_x$ antinodes. In the unfiltered modes, fluctuations are often much stronger in the core than in the shear layer. That the global and SPOD modes share this feature shows, dynamically and statistically, that trapped waves and shear-layer KH waves are almost un(cor)related. This is in line with the absence of saddle points linking trapped and KH local eigenmodes \citep{TowneEtAl2017JFM,MartiniEtAl2019JFM,PrinjaEtAl2025TCFD}. Exceptions are the symmetric schlieren SPOD modes in figure \ref{fig:cmp_modes_m0}(a--d), where core and shear-layer fluctuations may be conflated by line-of-sight integration. The global modes for larger $n_x$ do fluctuate more in the shear layer, explaining why, as a hallmark of shear flows \citep{TrefethenEtAl1993Science}, non-normality also grows (figure \ref{fig:pseudospectra}).

\begin{figure}
\centering
\begin{tikzpicture}[font=\scriptsize]
\node (graphics) {\includegraphics{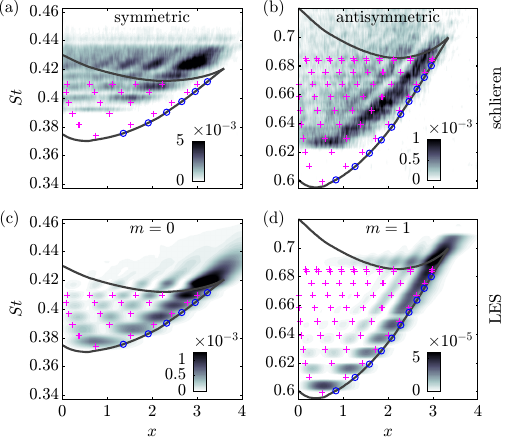}};
\node[darkgray] at (-1.8,1.4) {S1}; \node[darkgray] at (-2.8,2.8) {S2}; 
\node[darkgray] at (-1.8,-2.15) {S1}; \node[darkgray] at (-2.8,-.75) {S2}; 
\node[darkgray] at (3,2.1) {S1}; \node[darkgray] at (1.8,3.1) {S2}; 
\node[darkgray] at (3,-1.45) {S1}; \node[darkgray] at (1.8,-.45) {S2}; 
\end{tikzpicture}
\caption{Squared modulus of the $k\le0$ part of schlieren (a,b) and LES (c,d) SPOD modes, scaled by their eigenvalues, along $r=0$ for the symmetric or $m=0$ component (a,c) and $r=0.25$ for the antisymmetric or $m=1$ component (b,d). Panels (a) and (b--d) respectively show ranks one and two. Pluses mark antinodes of trapped global modes. Circles denote standing-wave frequencies from \eqref{subeq:standingWaveInPhase}. Curves are saddles.}
\label{fig:freq_x_SCHLES}
\end{figure}
Figure \ref{fig:freq_x_SCHLES} condenses the $k\le0$ part of all trapped SPOD and linear modes into the $St$--$x$ plane. Standing-wave frequencies are marked on the saddles only for reference. The SPOD and global modes are at least loosely bounded in $x$ by $x=0$ and the S1 curve. Between these limits, the modes are discrete in both $St$ and $x$. They form staggered antinodes, telltale signs of standing-wave resonance. We also note two slight differences between the models and the data, besides the same frequency shifts that figures \ref{fig:cmp_modes_m0}--\ref{fig:cmp_modes_m1} reported.
First, in figure \ref{fig:freq_x_SCHLES}(b), the S1 curve and the global modes' rightmost antinodes do not follow the slope of the darkest band in the schlieren contours. Second, in every panel, the global modes' leftmost antinodes are closer than the SPOD modes' to $x=0$. This may be due to a small impedance-mismatch between the inlet sponge of the linear solver and the nozzle in the experiment and LES.

\begin{figure}
\centering
\begin{tikzpicture}[font=\scriptsize]
\node (graphics) {\includegraphics{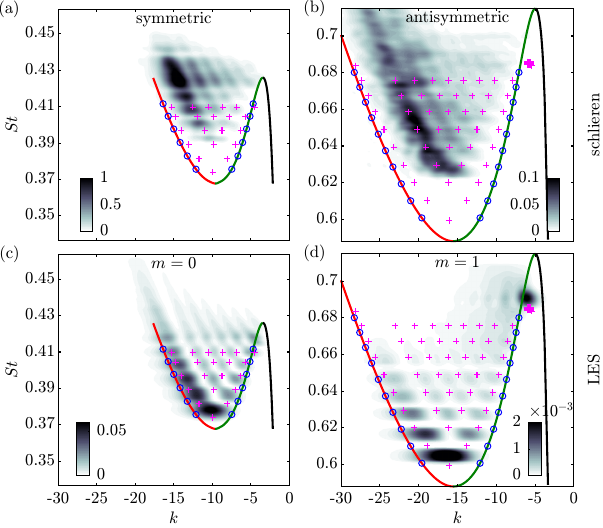}};
\node[red] at (-2.4,2) {$k_d^-$}; \node[soberGreen] at (-.75,2) {$k_T^+$}; \node at (-.35,2.8) {$k_p^-$}; 
\node[red] at (-2.4,-2.2) {$k_d^-$}; \node[soberGreen] at (-.75,-2.2) {$k_T^+$}; \node at (-.35,-1.4) {$k_p^-$}; 
\node[red] at (1,2.2) {$k_d^-$}; \node[soberGreen] at (3.8,2.2) {$k_T^+$}; \node at (4.4,2.2) {$k_p^-$}; 
\node[red] at (1,-2) {$k_d^-$}; \node[soberGreen] at (3.8,-2) {$k_T^+$}; \node at (4.4,-2) {$k_p^-$}; 
\end{tikzpicture}
\caption{Streamwise Fourier transform of figure \ref{fig:freq_x_SCHLES}. Curves are vortex-sheet dispersion relations. Circles are standing waves from \eqref{subeq:standingWaveInPhase}.}
\label{fig:freq_kx_SCHLES}
\end{figure}
The streamwise Fourier transform of SPOD and global modes over $0\le x\le4$ (zero-padded to $x=100$ to sharpen wavenumber resolution) yields the $St$--$k$ diagrams, or empirical dispersion relations, in figure \ref{fig:freq_kx_SCHLES}. Now the modes are loosely bounded in $k$ by the vortex-sheet duct-like waves, $k_d^-$ and $k_T^+$. Meanwhile, they exclude the acoustic jet wave, $k_p^-$. As such, trapped-mode resonance draws mainly from the wavenumber range $k_d^- \lesssim k \lesssim k_T^+$. Within this range, the modes are discrete in $St$ and $k$. But while most of them have $n_x$ antinodes per frequency, one stands out:
for $m=1$ in figure \ref{fig:freq_kx_SCHLES}(d), at $St\approx0.69$, the LES SPOD and global modes peak solely at $k\approx k_T^+$. Compact in $k$, the same modes are dispersed in $x$ (figure \ref{fig:freq_x_SCHLES}d), peaking where the S1 and S2 saddles meet and stretching towards the nozzle. For the schlieren SPOD, this structure is in rank one, so does not show up in the rank-two modes in figures \ref{fig:freq_x_SCHLES}(b) or \ref{fig:freq_kx_SCHLES}(b). We will speculate about the structure's origin in \S \ref{sec:discussion}.

The grid-like antinodes in the $St$--$x$ plane (figure \ref{fig:freq_x_SCHLES}) mirror those in the $St$--$k$ plane (figure \ref{fig:freq_kx_SCHLES}). That mirror relation reflects the duality between a Fourier transform pair.
In physical space, turning points at $x=0$ and S1 serve as boundary conditions that admit discrete solutions. As figure \ref{fig:freq_kx_SCHLES} shows, this principle holds too in wavenumber space, but with the parallel-flow dispersion relation playing the role of boundary conditions. In other words, while \eqref{subeq:standingWaveInPhase} implies that we prescribe $L$ and solve for discrete $\Delta k$, we can equivalently prescribe $\Delta k$ and solve for discrete $L$.

\section{Discussion}\label{sec:discussion}
Section \ref{sec:results} found excellent agreement among the experiment, the LES, and the models, leaving two details that are worth discussing. First, not all trapped modes line up precisely in frequency. Mode frequencies in the LES and schlieren data disagree by $0.01\lesssim\Delta St\lesssim0.02$. Although small, the gap points out that the LES is a model, not a replica, of the experiment. These frequencies (especially the schlieren's), in turn, are modestly above what linear theories predict. Maximising accuracy is not our goal; physical insight is. But as \S \ref{subsubsec:frankensteinTheory} noted, the linear models may benefit from stricter treatments of the turning points: weakly non-parallel stability from asymptotically matching two WKBJ solutions \citep{BenderOrszag1999Springer}, and global stability from including the nozzle. Moreover, linear stability around the experimental mean \citep[e.g.]{SuzukiColonius2006JFM}, rather than the simulated mean, is more likely to be faithful to the experiment.

Second, in figure \ref{fig:freq_kx_SCHLES}(d) the SPOD mode at $St\approx0.69$ is energetic and, unlike those below it, compact in $k$. It is approximated by a global mode (bold $\boldsymbol+$) in a non-normal part of the pseudospectrum (figure \ref{fig:pseudospectra}b), and therefore rooted in modal and non-modal global linear stability. That said, its distinctive structure invites local interpretations too. One hypothesis is that, at $St\approx0.69$, $k_p^-$ replaces $k_d^-$ as the upstream wave in resonance \citep{TowneEtAl2017JFM,PrinjaEtAl2025TCFD}. It is backed up by figure \ref{fig:freq_kx_SCHLES}(d), where the SPOD mode vanishes near $k_d^-$. But it is undermined by figure \ref{fig:freq_x_SCHLES}(d), where the same mode partly sits above the S2 curve. There, $k_p^-$ and $k_T^+$ are evanescent.
Another hypothesis is that long-range resonance between $k_d^-$ and $k_T^+$ switches to short-range resonance \citep{Bogey2025JFM}, i.e., a wavemaker or pocket of absolute instability \citep{ChomazEtAl1988PRL,HuerreMonkewitz1990AnnuRev}.
This also seems to fall short, as trapped waves are absolutely stable or absolutely neutral \citep{TowneEtAl2017JFM}. So at first glance, neither long-range resonance ($k_p^-$ with $k_T^+$) nor short-range resonance fits the $St\approx0.69$ mode. However, in figure \ref{fig:freq_x_SCHLES}, the frequency offsets between theories and data blur the boundary between propagation and evanescence. Additionally, evanescent, stable, or neutral waves may be destabilised by turbulent forcing. Also, the turning-point regions, where WKBJ solutions break down, allow for more intricate dynamics \citep{BenderOrszag1999Springer,MonkewitzEtAl1993JFM}. In particular, the S1 and S2 saddles meet at a `super branch point' \citep{ChesterEtAl1957CPS,Davies1989PoF,Healey2004JFM}, roughly where the SPOD- and global-mode amplitudes peak. On balance, the complexity of such analyses is a strong motivation for global analysis.

\section{Conclusions}\label{sec:conclusion}
A high-subsonic turbulent jet traps acoustic modes in its potential core, restricting them to branches of the dispersion relation. The branches each span a frequency band whose width is $\Delta St\sim O(0.1)$. \citet{SchmidtEtAl2017JFM} and \citet{TowneEtAl2017JFM} suggested that each band consists of $O(10)$ discrete frequencies, which at the time were too close to be distinguished in data. This experimental--numerical--theoretical effort confirms their hypothesis. To do so, we gather $O(100\,000)$ snapshots from dedicated schlieren experiments and LES. We apply an SPOD algorithm that uses the sine-taper estimator and the jet's two azimuthal symmetries, rotational and reflectional. For both data, SPOD resolves the discreteness. In the SPOD spectra, trapped modes form arcs of eigenvalues across ranks and frequencies. Neighbouring arcs, representing distinct structures, are $\Delta St\sim O(0.01)$ apart.
SPOD and global theory broadly agree on trapped modes' frequencies and structures. They find trapped and shear-layer modes to be (cor)related only weakly. In global stability spectra and $\varepsilon$-pseudospectra, the lowest-frequency trapped modes in a branch are the least modally-damped, while the highest-frequency ones are the most non-normal. Linear theory thus signals a contest between modal and non-modal stability. This is confirmed by SPOD, which shows that both are needed to explain the trapped modes' energy in the data.
The trapped SPOD and global modes are loosely bounded in the $St$--$x$ plane by $x=0$ and the S1 saddle points. In the $St$--$k$ plane, the bounds are the duct-like $k_d^-$ and $k_T^+$ waves. Along $x$ or $k$, the $n_x$th trapped mode has $n_x$ antinodes. The bounds, $x=0$ and S1 alongside $k_d^-$ and $k_T^+$, quantise trapped modes. An in-phase standing-wave resonance model encapsulates this principle, and predicts the trapped global modes' frequencies.

\backsection[Acknowledgements]{We are indebted to A. Lebedev and D. Eysseric for their experimental prowess. B.Y. has benefitted from L. Heidt's help with stability analysis. P.J and M.N.S are grateful to Airbus for their support throughout this work.
}

\backsection[Funding]{O.T.S. and B.Y. gratefully acknowledge support from Office of Naval Research grant N00014-23-1-2457, under the supervision of Dr. S. Martens. P.J and M.N.S were supported by the French government program ``France 2030'' (LABEX INTERACTIFS, reference ANR-11-LABX-0017-01), by the DGAC (\textit{Direction Générale de l’Aviation Civile}), by the PNRR (\textit{Plan National de Relance et de Résilience Français}) and by NextGeneration EU via the project MAMBO (\textit{Méthodes Avancées pour la Modélisation du Bruit moteur et aviOn}).
}

\backsection[Competing interests]{The authors declare none.}

\appendix

\section{Confidence in SPOD estimates}\label{sec:appConfidence}
\begin{figure}
\centering
\includegraphics{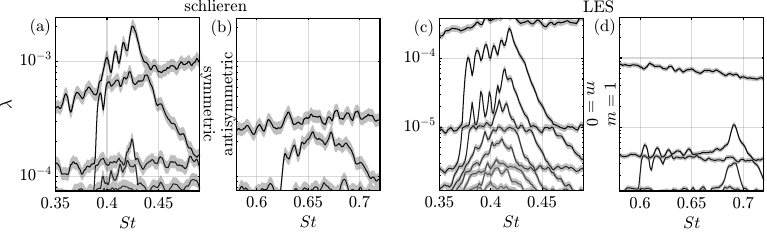}
\caption{SPOD spectra from figure \ref{fig:cmp_spectra} and their 95\% confidence intervals \eqref{eq:confidenceInterval} in grey.}
\label{fig:confidenceInterval}
\end{figure}
From the SPOD spectra in figure \ref{fig:cmp_spectra}, we concluded that trapped modes in the data are discrete. That conclusion rests on well converged SPOD estimates. This appendix quantifies their convergence using confidence intervals around the eigenvalues. We follow standard practice \citep{PercivalWalden2020Cambridge,SchmidtColonius2020AIAAJ} and assume the eigenvalue estimates are distributed roughly as $\chi^2_\nu$ with $\nu$ degrees of freedom. The estimator in \S \ref{subsec:SPOD}, which segments the data into $\nblk$ blocks, tapers each block by $\nwin$ windows, and weights the $n$th window by $\mu_n = 6[\nwin^2-(n-1)^2]/(4\nwin^3+3\nwin^2-\nwin)$ \citep{RiedelSidorenko1995IEEE,YeungSchmidt2024TCFD}, has $\nu = 2\nblk/\sum_{n=1}^{\nwin} \mu_n^2$ degrees of freedom. Because of reflectional symmetry, for only $m\neq0$ in the LES, $\nu = 4\nblk/\sum_{n=1}^{\nwin} \mu_n^2$. Defining $Q_\nu(\alpha)$ as the point on $\chi^2_\nu$ at which the probability $P[\chi^2_\nu \le Q_\nu(\alpha)] = \alpha$, we get the $(1-2\alpha)100\%$ confidence interval
\begin{equation}
\qty[{\nu\lambda}/{Q_\nu(1-\alpha)}, {\nu\lambda}/{Q_\nu(\alpha)}]. \label{eq:confidenceInterval}
\end{equation}
Figure \ref{fig:confidenceInterval} overlays our SPOD eigenvalues in the $n_r=1$ frequency band with their 95\% confidence intervals. For the LES, the intervals are narrow, wrapping tightly around the trapped modes' peaks. These estimates are thus likely to be close to the true eigenvalues. For the schlieren, relative to the height of the peaks, the intervals are wider. In the symmetric component, the peaks at $St\approx0.4$ and $0.425$ remain clear, while the other peaks are less so. We cannot distinguish the wiggles in the antisymmetric component from noise. But figures \ref{fig:cmp_modes_m0}--\ref{fig:freq_kx_SCHLES} also showed the $n_x$th peak has a mode with $n_x$ antinodes, linking the discreteness in frequency to that in space. The totality of the experimental and numerical evidence affirms that trapped modes are spatiotemporally quantised.

\bibliographystyle{plainnat}
\bibliography{refs}

\end{document}